# On operational approach to entanglement and how to certify it

M. Kupczynski

Département de l'Informatique, UQO, Case postale 1250,
succursale Hull, Gatineau. Québec, Canada J8X 3X 7

Entangled physical systems are an important resource in quantum information. Some authors claim that in fact all quantum states are entangled. In this paper we show that this claim is incorrect and we discuss in operational way differences existing between separable and entangled states. A sufficient condition for entanglement is the violation of Bell- CHSH-CH inequalities and/or steering inequalities. Since there exist experiments outside the domain of quantum physics violating these inequalities therefore in the operational approach one cannot say that the entanglement is an exclusive quantum phenomenon. We also explain that an unambiguous experimental certification of the entanglement is a difficult task because classical statistical significance tests may not be trusted if sample homogeneity cannot be tested or is not tested carefully enough.

## 1. Introduction

The entanglement of quantum states is considered to be an important resource in Quantum Information therefore it should be well understood.

According to a common interpretation of quantum theory (QT) individual results of measurements are obtained in irreducibly random way. Therefore if in a laboratory *x* a physical observable *A* is measured on identically locally prepared physical systems I and in a distant laboratory *y* a physical observable *B* is measured on identically locally prepared systems II the results of these measurements should be independent.

Using the language of mathematical statistics we have here two random experiments measuring values of two independent random variables *A* and *B* such that their joint probability distribution factorizes and their expectation values $E(A,B)= E(A)E(B)$ or in other words their covariance $cov(A,B) =0$. In QT a quantum state of a "pair I+II" is separable and is described by a tensor product of quantum states of systems I and II.

In 1935 Einstein, Podolsky and Rosen[1] demonstrated that results of local measurements on two quantum systems I and II which interacted in the past and separated afterwards were strongly correlated and had to be described by a particular non-separable state which was called entangled by Schrödinger[2].

If two or several systems do interact forming a multi-partite system then various degrees of freedom of this system can be strongly correlated and the system is in so called generalised entangled state in contrast to the entangled state for distant non- interacting systems. It was noticed by Zanardi et al.[3,4] that a Hilbert space of possible state vectors of these physical systems could be partitioned by introducing various tensor product structures (TPS) induced by experimentally accessible observables (interactions and measurements). Similar conclusions were reached by Barnum et al.[5]. In this sense the entanglement is relative to a particular set of experimental capabilities.

Continuing this line of thought Torre et al.[6] claimed to prove that any separable state was in fact entangled. They used the non-vanishing of quantum covariance function (QCF) as a criterion for the entanglement. Starting from separable quantum states and uncorrelated local observables *A* and *B* they showed that it was easy to find two functions *F (A,B)* and *G(A,B)* such that their QCF was non-zero and concluded that the entanglement was a universal property of any quantum multipartite state. This point of view was shared in several references[7, 8].

In section 2 after defining in operational way separable and entangled states we recall that a non-vanishing of covariance between some observables is not a sufficient condition for proving the entanglement. The entanglement can only be proven, if having sufficient experimental capabilities, one may demonstrate the violation of Bell and CHSH inequalities (BI-CHSH) and /or steering inequalities [9-11]. The classification of entangled states is not easy because there exist entangled mixed quantum states which do not violate BI-CHSH or steering inequalities [12]. For more complicated multipartite systems one has to use so called entanglement witnesses [13, 14].

In section 3 we analyze the examples from Torre et al.[6] and show that functions *F(A,B)* and *G(A,B)*, they construct, do not correspond to any new directly measurable observables. Their values can be only calculated using the observed values of uncorrelated observables *A* and *B*. Therefore we may say that the experimental capabilities [3-5] do not allow a new TPS in the Hilbert space induced by *F* and *G*. Since *F(A,B)* and *G(A,B)* are in general dependent random variables therefore the non-vanishing of their covariance is obvious and by no means can be considered as a proof that a separable quantum state is in fact entangled.

In section 4 in order to make clear that the non- vanishing of covariance has nothing to do with the entanglement we give an example of local random experiments with pairs of fair dices in distant locations giving correlated outcomes.

In section 5 we review very shortly some results obtained in collaboration with Hans De Raedt [15]. We demonstrated that sample inhomogeneity might invalidate in a dramatic way standard statistical significance tests. If sample homogeneity is not studied, data suffer from *sample homogeneity loophole* [16] and the certification of entanglement cannot be fully trusted. We do not doubt that Bell-CHSH-CH inequalities are violated in spin polarization correlation experiments (SPCE) [17-20] but we point out that in future experiments sample homogeneity has to be checked carefully in order to assure unambiguous results.

Section 6 contains our conclusions.

2. **Operational definition of the entanglement**

Let us consider an ensemble of identically prepared pairs of physical systems on which we can perform coincidence measurements of some local physical observables *A* and *B*. Our ensemble can be two beams of "particles" sent by some source *S* or an ensemble of pairs of "quantum dots" obtained by resetting of a state of a particular pair of "quantum dots" before each repetition of local measurements etc.

The outcomes of local experiments have in general a statistical scatter thus can be interpreted as results of measurements of some random variables *A* and *B* for which we may define the expectation values and the covariance. If we use QT to describe our experiments we introduce a state vector $\psi \in H_1 \otimes H_2$ or a density operator $\rho$. In a discrete case local observables are represented by Hermitian operators $\hat{A}_1 = \hat{A} \otimes I$ and $\hat{B}_1 = I \otimes \hat{B}$. Using this notation we define conditional expectation values: $E(A|\psi) = \langle \psi, \hat{A}_1 \psi \rangle$ or $E(A|\rho) = Tr \rho \hat{A}_1$ etc. We will use in the following the second form which is more general because it applies also to mixed quantum ensembles.

A conditional covariance of *A* and *B* is defined by:

$$\text{cov}(A, B | \rho) = E(AB | \rho) - E(A | \rho) E(B | \rho) \tag{1}$$

The conditional covariance of *A* and *B* coincides with QCF used by Torre et al.[6]. If $\rho$ is a separable state and random variables *A* and *B* are independent (corresponding Hermitian operators commute) then the conditional covariance function defined by Eq.1 has the obvious property:

$$\text{cov}(kA + nB, mA + lB | \rho) = km \, \text{var}(A | \rho) + nl \, \text{var}(B | \rho) \tag{2}$$

Using the above introduced notation we may define separable, non-separable and entangled quantum states of two distant physical systems:

A state $\rho$ is *separable* if it can be written in the form $\rho = \rho_1 \otimes \rho_2$ where $\rho_1$ and $\rho_2$ are density operators acting in the Hilbert spaces $H_1$ and $H_2$ respectively. For separable states $\text{cov}(A, B | \rho)$ vanishes for all measurable pairs of local observables (*A*, *B*).

A state $\rho$ is a *convex sum of separable states* if $\rho = \sum_{i=1}^{k} p_i \rho_i \otimes \tilde{\rho}_i$ with $0 < p_i < 1$. Conditional expectation values for all pairs of local observables (*A*, *B*) can be written now as:

$$E(AB | \rho) = \sum_{i=1}^{k} p_i E(A | \rho_i) E(B | \tilde{\rho}_i) \tag{3}$$

and $\text{cov}(A, B | \rho)$ does not vanish if some of products of local expectation values are different from zero. Besides if $|E(A | \rho_i)| \leq 1$ and $|E(B | \tilde{\rho}_i)| \leq 1$ one can easily prove CHSH inequalities:

$$|E(AB | \rho) - E(AB' | \rho)| + |E(A'B | \rho) + E(A'B' | \rho)| \leq 2 \tag{4}$$

Therefore the correlations for a convex sum of separable states can be reproduced by so called local stochastic hidden variable (SHV) models [9, 35, 36].

A state $\rho$ is *entangled* if it cannot be written as a *convex sum of separable states*. It can be checked using quantum state tomography, entanglement witnesses or by showing that expectation values for some pairs of measurable local observables violate Bell-CHSH or steering inequalities.

Since a probabilistic SHV model is particularly suited to describe a convex sum of separable states it is not strange that it does not provide a correct probabilistic model for the experiments with entangled states.

3. **Not all quantum states are entangled**

In their first example Torre et al.[6] consider a quantum system described by a separable quantum state $\psi$ depending on position coordinates. The local position observables $X_1$ and $X_2$ are represented by $\hat{X}_1 = \hat{X} \otimes I$ and $\hat{X}_2 = I \otimes \hat{X}$ where $\hat{X}$ is a position operator. Of course $[\hat{X}_1, \hat{X}_2] = 0$ and $cov(X_1, X_2 | \psi) = 0$.

Next they introduce two other variables $A = X_1 + X_2$ and $B = X_1 - X_2$. Using a similar formula to one given in Eq.2 they prove that $cov(A, B | \psi)$ is equal to the difference of conditional variances of $A$ and $B$ which is in general different from zero and they conclude that the state $\psi$ is in fact entangled. Since it is impossible to measure the observables $A$ and $B$ directly on two systems prepared in the state $\psi$ therefore these observables do not introduce a new TPS in the Hilbert space of states and the state $\psi$ does not become entangled.

The same argument applies to their second example in which they consider two free spin $\frac{1}{2}$ particles in another separable state $\psi$. They define new observables $S_z^2 = (S_z \otimes I + I \otimes S_z)^2$ and $S_x^2 = (S_x \otimes I + I \otimes S_x)^2$ prove that the conditional covariance does not vanish. Since the values of these observables can be only deduced from local measurements of the spin projections $S_z$ and $S_x$ thus the initial separable quantum state does not become entangled.

The notion of entanglement was generalized in order to describe the coupling of different degrees of freedom of a single compound quantum system [4,5] for example the hydrogen atom and is called a generalized entanglement.

In this approach a quantum state of a system is called "entangled" if it is not a convex sum of tensor products of some vectors representing different degrees of freedom of the compound system. For example for the hydrogen atom we find a natural factorization *CM-R* where *CM* denotes a center- of -mass and *R* relative motion degrees of freedom. Thus in *CM-R* splitting the quantum state of hydrogen is "separable".

If we use instead the degrees of freedom of proton and electron, so called *e-p* splitting, then formally the state vector of hydrogen atom is "entangled" in these new degrees of freedom. However the "center-of -mass+ relative degrees of freedom" structure appears as primarily operable form of the experimental reality of atoms [7]. The e-p splitting structure becomes only useful when the atom ionizes.

Let us note that talking about a "separable" quantum state in case of CM-R splitting of a state vector of an atom takes us far from the original idea of a separable state describing two non-interacting distant physical systems [1,2] and may easily lead to confusion. If we stick to the operational definition of separable and entangled states we stay on a safe ground.

## 4. "Non- separable" ensemble of pairs of fair dices

Let us consider a following probabilistic random experiment.

Carol can send pairs of dices to Alice and Bob. She has two types of fair dices D1 with 1 written on three faces and 0 on the remaining three faces and D2 with 1 written on 4 faces and 0 on the remaining two faces. She chooses to send a pair (D1, D1) with a ``probability`` 0.25 and a pair (D2, D2) with a ``probability`` 0.75. Alice and Bob roll received dices and record their observations 0 or 1 and compare them. Using the language of mathematical statistics they measure the values of the corresponding random variables $A$ and $B$ on some mixed classical statistical ensemble. It is easy to see that

$$E(A) = E(B) = \frac{1}{4} \times \frac{1}{2} + \frac{3}{4} \times \frac{2}{3} = \frac{5}{8} \quad E(A,B) = \frac{1}{4} \times \frac{1}{2} \times \frac{1}{2} + \frac{3}{4} \times \frac{2}{3} \times \frac{2}{3} = \frac{19}{48} \text{ thus cov}(A,B) \neq 0.$$

If we used a "quantum like" model [30] we would conclude that our "quantum like" state is not separable but of course not entangled. We could introduce new classical dependent random variables $F(A,B)$ and $G(A,B)$ without gaining any new information about the "quantum like "state of the ensemble of dices sent by Carol to Alice and Bob.

## 5. Certification of entanglement and sample homogeneity loophole

As we mentioned in introduction a certification of entanglement requires testing of various Bell-type inequalities. We have shown recently with Hans de Raedt that significance tests might break down dramatically if a studied sample was not homogeneous[15]. One may not assume that experimental runs produce 'simple' random samples without verifying it. In particular a careful study of sample homogeneity has to be incorporated in experiments testing Bell-CHSH-CH inequalities.[17-20]

The standard classical inference: significance tests and asymptotic theorems are based on the assumption that data sets are *simple random samples*. According to standard sampling methods a sample $S=\{x_1,x_2,…x_N\}$ of size N is interpreted as an observation of a multivariate random variable $\{A_1,A_2,…A_N\}$.

S is a *simple random sample* if and only if:

- all trials are independent which means: all $A_i$ are independent random variables
- S is homogenous which means: all $A_i$ are identically distributed random variables .

We studied a random experiment in which a signal was entering a measuring device and from time to time some discrete outcomes were produced and a sample S was obtained. We assumed a simple probabilistic model of this random experiment:

- a signal is described by a probability distribution $p_1(m)$
- a state of the device at the moment of a measurement is described by a probability distribution $p_2(n)$
- an output of the device is one of the discrete values $A(m, n)$.

If this simple model is assumed then

$$\langle A \rangle = \sum_{m,n} A(m,n) p_1(m) p_2(n) \qquad (4)$$

and p(A(m, n)=a) and the standard deviation $\sigma_A$ are easily found.

Probabilistic model does not give the information about a detailed internal protocol which is used by the device to output successive outcomes. Therefore we perform several Monte Carlo simulations using various possible internal protocols and we compare properties of finite samples generated by these protocols. One family of protocols we call ($N_1$, $N_2$, n) :

- generate one value of *m* and $N_2 > 1$ values of *n* using $p_1(m)$ and $p_2(n)$
- evaluate *A(m, n)* and output the values for the $N_2$ different values of *n*
- repeat the process $N_1$ times in order to create a sample of a size N=$N_1$ $N_2$

We created samples containing $10^3$ and $10^6$ data items by choosing $N_1$=4 or 40 and $N_2$=250, 2500 and 25000 or vice-versa. By repeating the computer experiments 100 times, we generated large random samples containing even $10^8$ outcomes subdivided into 100 bins. We have checked that our conclusions did not depend on the particular random number generator used and that they did not change when we repeated the experiments.

In the limit when both $N_1$ and $N_2$ tend to infinity one might expect that estimates of proportions and averages for all different protocols should be consistent but for our large finite samples it was not true

In particular we tested an inequality: Test $H_0$: $\langle A \rangle_S / \langle A \rangle \leq 1$ where $\langle A \rangle_S$ was a sample average and $\langle A \rangle$ a theoretical expectation value found using a specific probabilistic model (4).

Using a protocol (4, 25000, n) we generated 3 samples containing $10^5$ data items for which we could confidently reject the hypothesis $H_0$ because the inequality was violated by more than 2000 SEM (standard errors of the mean). However when we generated 100 samples creating a huge sample containing $10^7$ data items the average was 0.9727 thus the inequality was satisfied and we could not reject the hypothesis $H_0$. The reason was that our huge sample was not homogeneous what we confirmed by additional homogeneity tests we performed.

The violation of various inequalities was reported in many experiments but often only a few large data samples were studied. For example in Weihs et al. experiment two long runs were studied and only in one of them CHSH inequality was significantly violated.[17] In Giustina et al. experiment only one long run subdivided into 30 bins was studied.[19] The sample homogeneity was not tested carefully enough or could not be tested .[16]

Several authors pointed out that various proofs of Bell type inequalities use a counterfactual reasoning and suffer from a *contextuality loophole* .[21-38] Therefore we do not doubt that Bell type inequalities can be and are violated in SPCE but the significance of the violation should be confirmed by additional homogeneity tests if it is possible.[15,16]

## 6. Conclusions

In the operational approach the entanglement is an objective property of an ensemble of various expectation values *E(A,B)* found in coincidence experiments performed on some identically prepared physical systems in different experimental settings.

A criterion of non-vanishing QCF= $\text{cov}(A,B\,|\,\rho)$ for some physical observables used by Torre et al. [6] in order to conclude that quantum states are entangled is insufficient.

To prove whether physical systems are prepared in an entangled quantum state one has to show that not all correlations between available local variables can be explained by using a convex sum of separable quantum states.

This can be proven if BI-CHSH or steering inequalities are violated. Due to ambiguities related to the finite statistics, efficiency of detectors, widths of coincidence windows, post selection, noise etc. it is a difficult but not an impossible task to accomplish.[11, 17-20]

However the significance of these results has to be confirmed by additional sample homogeneity tests because as we showed a *sample homogeneity loophole* can invalidate significance tests.[15, 16]

Since there exist many random experiments from outside the domain of quantum physics in which BI-CHSH are violated [29, 30] therefore the entanglement as defined operationally is not exclusively a quantum phenomenon.

It is not strange since BI-CHSH inequalities may be interpreted as necessary conditions for the existence of joint probability distributions of values of several dichotomous random variables which can be measured pair-wise but not simultaneously.[22,23, 27-30, 33]

It is a not well known but it is possible to simulate, in a consistent and local way, many experiments from quantum optics and neutron interferometry [38-40] including those violating BI-CHSH inequalities.

Quantum properties of entangled signals led to many successful applications in quantum cryptography. However the feasibility of scalable quantum computer depends not only on technological progress [43] but also on understanding of foundations of quantum theory [24-26, 29, 41-46] and a subtle notion of probability.

## Acknowledgements

I am grateful to UQO for a travel grant and to Andrei Khrennikov for the invitation to give a talk during this interesting QTFT conference and for his kind hospitality. I would like to thank also the anonymous referees of this paper for several valuable suggestions.

## References

1. A. Einstein, A. Podolsky and N. Rosen, *Phys. Rev.* **47** (1935) 777.
2. E. Schroödinger, *Proc. Cambridge.Philos.Soc.* **31** (1936) 553.
3. P. Zanardi, *Phys.Rev.Lett.*, **87** (2001) 077901.
4. P. Zanardi, A. Lidar and S. Lloyd, *Phys.Rev.Lett.* **92** (2004) 060402.
5. H. Barnum H. et al, *Phys.Rev.Lett.* **92** (2004) 107902.
6. A.C. de la Torre, D. Goyeneche and L. Leitao, *Eur.J.Phys.* **31** (2010) 325.

7. J. Jeknic-Dugic, M. Dugic, A. Francom, and M. Arsenijevic, *OAlib* (2014) doi: [10.4236/oalib.1100501](10.4236/oalib.1100501).
8. M. Caponigro and E. Giannetto, *Discusiones Filosóficas* **13** (2012) 137.
9. J.S. Bell, *Speakable and Unspeakable in Quantum Mechanics*, (Cambridge UP, Cambridge, 2004).
10. H.M. Wiseman, S.J. Jones and A. Doherty, *Phys.Rev.Lett.* **98** (2007) 140402.
11. B. Wittmann, et.al., *New J. Phys.*, **14** (2012) 053030.
12. R.F. Werner, *Phys. Rev.* **A 40** (1989) 4277.
13. P. Hyllus, O. Gühne, D. Bruß, and M. Lewenstein, *Phys. Rev. A* 7**2** (2005) 012321.
14. M. Kotowski, M., Kotowski and M. Kuś, *Phys. Rev. A* **81** (2010) 062318.
15. M. Kupczynski, and H. De Raedt, *Comp. Phys. Comm.* (2015) doi: 10.1016/j.cpc.2015.11.010)
16. M. Kupczynski, quant-Ph/1505.063449.
17. G. Weihs, et al., *Phys. Rev. Lett.* **81** (1998) 5039.
18. M.A. Rowe, et al., *Nature* **409** (2012) 791.
19. Giustina M. et al., *Nature* **497** (2013) 227.
20. B.G. Christensen, et al., *Phys. Rev. Lett.* **111** (2013) 1304.
21. M. Kupczynski, *Phys.Lett. A* **121** (1987) 205.
22. N. Vorob'ev, *Theor. Probab. Appl.* 7, (1962) 147.
23. I. Pitovsky, *Brit.J.Phil.Sci.* **45** (1994) 95.
24. A. Yu. Khrennikov, *Interpretation of Probability*, (VSP, Utrecht, 1999)
25. M. Kupczynski, *AIP Conf. Proc.* **861** (2006) 516.
26. M. Kupczynski, *AIP Conf. Proc.* **962** (2007) 274.
27. A. Yu. Khrennikov, *AIP Conf. Proc.* **962** (2007) 121.
28. A. Yu. Khrennikov, *AIP Conf. Proc.* **1101** (2009) 86.
29. A. Yu. Khrennikov, *Contextual Approach to Quantum Formalism*, (Springer, Dortrecht, 2009)
30. A. Yu. Khrennikov, *Ubiquitous Quantum Structure*, (Springer, Berlin, 2010).
31. T.M. Nieuwenhuizen, *AIP Conf. Proc.* **1101** (2009) 127.
32. T.M. Nieuwenhuizen, *Found. Phys.*, (2010), doi: 10.1007/s10701-010-9461-z.
33. K. Hess, H. De Raedt, K. Michielsen, *Phys. Scr.* **T151** (2012) 014002.
34. M. Kupczynski, *AIP Conf. Proc.* **1508** (2012) 253.
35. M. Kupczynski, *J. Phys.: Conf. Ser.* **504** (2014) 012015. doi:10.1088/1742-6596/504/1/012015K
36. M. Kupczynski, *Found. Phys.* (12 Dec 2014), doi:10.1007/s10701-014-9863-4
37. A.Yu. Khrennikov, *Found.Phys.* **45** (2015) 711.
38. M. Żukowski, M and Č. Brukner, *J. Phys. A: Math. Theor.* **47** (2014) 424009.
39. H. De Raedt, K. Hess and K. Michielsen, *J. Comp. Theor. Nanosci.* **8** (2011) 1011.
40. K. Michielsen, F. Jin, and H. De Raedt, *J. Comp. Theor. Nanosci.* **8** (2011) 1052.
41. H. De Raedt H. and K. Michielsen, Ann.Phys.(Berlin) **524** (2012) 393
42. L.E. Ballentine, *Rev.Mod.Phys.* **42** (1970) 358.
43. L.E. Ballentine, *Quantum Mechanics: A Modern Development*, (World Scientific, Singapore, 1998).
44. M. Kupczynski, quant-Ph/0408002.
45. A.E. Allahverdyan, R. Balian and T.M. Nieuwenhuizen, *Physics Reports* 525 (2013) 1.
46. Kupczyski M. quant-Ph/1505.06348.